Fast automatic segmentation of thalamic nuclei from MP2RAGE acquisition at 7 Tesla

Running title: thalamic nuclei from MP2RAGE

Ritobrato Datta[1], Micky K Bacchus[1], Dushyant Kumar[2], Mark Elliott[2], Aditya Rao[3], Sudipto Dolui[2], Ravinder Reddy[2], Brenda L Banwell[1,4], Manojkumar Saranathan[5]

[1]Division of Neurology, Children's Hospital of Philadelphia, Philadelphia, PA;
[2]Departments of Radiology, Perelman School of Medicine at the University of Pennsylvania, Philadelphia, PA;
[3]Biological Basis of Behavior Program, University of Pennsylvania, Philadelphia, PA;
[4]Departments of Neurology, Perelman School of Medicine at the University of Pennsylvania, Philadelphia, PA;
[5]Department of Medical Imaging, University of Arizona, Tucson, AZ

Address correspondence to:

Ritobrato Datta, Ph.D.
email: dattar@email.chop.edu
Phone: 215-590-7291
ORCID: 0000-0003-1503-9381

ABSTRACT

Purpose: Thalamic nuclei are largely invisible in conventional MRI due to poor contrast. Thalamus Optimized Multi-Atlas Segmentation (THOMAS) provides automatic segmentation of 12 thalamic nuclei using white-matter-nulled (WMn) MPRAGE sequence at 7T. Application of THOMAS to Magnetization Prepared 2 Rapid Gradient Echo (MP2RAGE) sequence acquired at 7T has been investigated in this study.

Methods: 8 healthy volunteers and 5 pediatric-onset multiple sclerosis patients were recruited at the Children's Hospital of Philadelphia and scanned at Siemens 7T with WMn-MPRAGE and multi-echo MP2RAGE (ME-MP2RAGE) sequences. White-matter-nulled contrast was synthesized (MP2-SYN) from $T_1$ maps from ME-MP2RAGE sequence. Thalamic nuclei were segmented using THOMAS joint label fusion algorithm from WMn-MPRAGE and MP2-SYN datasets. THOMAS pipeline was modified to use majority voting to segment the bias corrected MP2-UNI images. Thalamic nuclei from MP2-SYN and MP2-UNI images were evaluated against corresponding nuclei obtained from WMn-MPRAGE images using dice coefficients, volume similarity indices (VSI) and distance between centroids.

Results: For MP2-SYN, dice > 0.85 and VSI > 0.95 was achieved for the 5 larger nuclei and dice > 0.6 and VSI > 0.7 was achieved for the 7 smaller nuclei. The dice and VSI were slightly higher whilst the distance between centroids were smaller for MP2-SYN compared to MP2-UNI, indicating improved performance using the synthesized WMn image.

Discussion: THOMAS algorithm can successfully segment thalamic nuclei in routinely acquired MP2RAGE images with essentially equivalent quality when evaluated against WMn-MPRAGE, hence has wider applicability in studies focused on thalamic involvement in aging and disease.

**Keywords**
Thalamic nuclei segmentation, multiple sclerosis; MP2RAGE, 7 Tesla

INTRODUCTION

The thalamus is composed of multiple functionally distinct nuclei, each having connections to multiple regions of the cortex, subcortical structures (basal ganglia, amygdala) and infratentorial structures (cerebellum and midbrain) (1,2). The thalamus acts as a 'relay center' or a 'gateway' of information processing and transmission throughout the brain (3,4). Reduction in whole thalamic volume, as well as changes in thalamic shape, have been reported in pediatric-onset and adult multiple sclerosis (MS) (5–7), Parkinson's (8), Alzheimer's disease (9), schizophrenia (10), alcohol use disorder (AUD) (11), Korsakoff's syndrome (12), and migraine (13). Changes at the level of specific thalamic nuclei have also been reported in MS and AUD (14–16). Thalamic nuclei are largely invisible in conventional MRI due to their small size and poor intra-thalamic contrast. Visualization of thalamic nuclei requires specialized MR sequences that enhance tissue contrast to delineate borders of individual thalamic nuclei. As a clinically relevant example, accurate visualization of thalamic nuclei aids the guidance of patient-tailored deep brain stimulation rather than using a stereotactic reference (17). MR identification of the specific thalamic nuclei to be targeted improves the accuracy and safety of focused ultrasound thalamotomy and reduces potential errors due to mis-targeting (18).

Task-based fMRI or tractography have been used to identify functional boundaries in the thalamus based on thalamo-cortical connections instead of anatomical distinctions (19–21). Diffusion-based methods have segmented the thalamus by using tensor models (22) and k-means clustering (23–26) or spherical harmonic decomposition of orientation distribution functions (27). Resting state fMRI-based methods have been used to segment the thalamic nuclei (28). However, lower spatial resolution of the underlying echoplanar imaging sequences and image distortion renders the segmentation of smaller thalamic nuclei inherently challenging.

At 7T, individual sequences have different coil heterogeneity profiles that require separate calibration sequences for bias-field correction. The Magnetization Prepared 2 Rapid Gradient Echo (MP2RAGE) sequence acquires two MPRAGE-like images at two different inversion times (TI) to generate bias-field corrected images (29). Compared to MPRAGE, MP2RAGE images yield better tissue contrast in the cortex and deep gray matter structures, that highlight anatomical boundaries of thalamic and midbrain nuclei. The intensity variations in these regions also correlate with myelin concentration (30,31). As a result, MP2RAGE is beginning to be routine in 7T neuroimaging protocols.

Recently, Thalamus Optimized Multi-atlas Segmentation (THOMAS) has been successfully used for automatic segmentation of 12 thalamic nuclei using a variant of the MPRAGE sequence called white-matter-nulled (WMn) MPRAGE (32,33). The automatic thalamic nuclei segmentation obtained from WMn-MPRAGE images demonstrated high accuracy, when validated by an expert against manual segmentation using the Morel histological atlas as a guide (34). However, it requires the acquisition of a special WMn-MPRAGE sequence which is not commonly accessible in clinical MRI scanners.

In this work, we investigated the use of THOMAS in conjunction with a multi-echo variant of MP2RAGE called ME-MP2RAGE (35) acquired at 7T for thalamic nuclei segmentation. We first evaluated the performance of THOMAS algorithm directly on the bias-corrected $T_1$-weighted image (MP2-UNI) obtained from post processing of the ME-MP2RAGE sequence. We also evaluated the performance of THOMAS on white-matter-nulled images synthesized from the quantitative $T_1$ map obtained from ME-MP2RAGE sequence. Both methods were compared against a separately acquired WMn-MPRAGE dataset on 13.

METHODS

*Participants*
8 healthy volunteers (20 yrs ± 2, 3 females) and 5 pediatric-onset MS patients (20 yrs ± 2, 2 females) were recruited. Healthy youth were recruited by local advertisement. Participants with relapsing-remitting MS, as defined by the 2017 McDonald criteria (37), and whose first attack occurred at less than 18 years of age were recruited from the Pediatric MS Program at the Children's Hospital of Philadelphia (CHOP). The study was approved by the CHOP Institutional Review Board, and informed written consent was obtained from all participants.

*7T MRI Protocol*
Participants were scanned at 7T (Siemens Magnetom Terra) at the University of Pennsylvania using a 32-channel head coil (Nova Medical) with the following sequences. WMn-MPRAGE parameters - resolution 1 mm isotropic, TR = 6000 ms, TE = 3.28 ms, TI = 670 ms, flip angle 5°, scan time ~ 7 mins. ME-MP2RAGE parameters: resolution 0.66 mm isotropic, TR = 6000 ms, TE1-4 = 1.91/4.02/6.13/8.24 s, TI1/TI2 = 750/2950 ms, scan time ~ 13 mins. ME-MP2RAGE is a variant of the conventional single-echo MP2RAGE with similar inversion times as the conventional MP2RAGE images but with four bipolar gradient readouts (38).

*Synthesized white-matter-nulled MP2RAGE data*
We synthesized white-matter-nulled contrast images from $T_1$ maps obtained from the ME-MP2RAGE sequence as per the following inversion recovery signal equation:

$$M_{synth} = M_0(1-2*exp(-TI/T_1))$$

where TI was set to 750 ms for suppressing white-matter and $T_1$ is the quantitative $T_1$ image obtained from the ME-MP2RAGE sequence. We refer to this synthesized dataset henceforth as MP2-SYN.
Figure 1 shows a representative axial slice from the raw ME-MP2RAGE data corresponding to the first and second inversion (a,b), the $T_1$ map (c) and the bias-corrected $T_1$ weighted image (d), the WMn contrast synthesized from the $T_1$ map (e) and lastly the independently acquired WMn-MPRAGE sequence (f) for comparison. The WMn contrast images (e-f) show improved intra-thalamic contrast as well as clear outer boundaries of the thalamus (yellow arrows) and the mammillothalamic tract, a small white matter structure (red arrows).

*Data Processing:*
All input images were preprocessed using the steps in the THOMAS workflow (39) and summarized here (Figure 2). A template image (created from mutual registration and averaging of 20 prior WMn-MPRAGE datasets (available as part of THOMAS source) (34) was first manually cropped using a bounding box restricted to the thalami and surrounding structures (Figure 2). After rigid registration of the template and the input image and warping the template crop mask, the input images were automatically cropped ensuring complete coverage of the left and right thalami. The cropped input image was diffeomorphically registered to the cropped template using ANTs (SyN option). The cropped priors were warped to the input image by a two-step warping by concatenating (a) the warps between the priors and the template (Figure 2, label $T_i$) and (b) the inverse warps between the template and the input image (Figure 2, labels $W^{-1}$ or $R^{-1}$ depending on the input image type).

*Thalamic Nuclei Segmentation*
The default THOMAS pipeline using joint label fusion was used to segment the thalamic nuclei on the WMn-MPRAGE images (Figure 2 left column) and on the MP2-SYN images (synthesized WMn images are similar in contrast to the true WMn-MPRAGE images). To apply THOMAS to the bias corrected $T_1$-weighted images (MP2-UNI) generated from the ME-MP2RAGE sequence, the THOMAS pipeline was modified to use majority voting instead of joint fusion to fuse the labels warped from priors. This was done as (a) the priors were labeled on WMn-MPRAGE sequence and (b) the contrast in the MP2-UNI images are closer to a $T_1$-weighted image. The modified THOMAS algorithm using majority voting is less dependent on the inherent contrast of the input anatomical image and relies on accuracy of registration of the priors and majority voting algorithm to achieve the final segmentation of the thalamic nuclei. On each subject, the mammillothalamic tract (MTT) and the following eleven nuclei were identified on the left and right thalami - medial group (mediodorsal nucleus, center median nucleus, habenula); posterior group (pulvinar, medial geniculate nucleus, lateral geniculate nucleus), lateral group (ventral posterolateral nucleus, ventral lateral anterior nucleus, ventral lateral posterior nucleus, ventral anterior nucleus), and anterior group (anteroventral nucleus).

*Segmentation Accuracy*
For each subject, the segmentations obtained on the independently acquired WMn-MPRAGE data was used as the ground truth. To compare the performance of the segmentations obtained on the images from the ME-MP2RAGE sequence, MP2-SYN images from a subject were first affine registered to that subject's WMn-MPRAGE image using ANTs (37). The thalamic labels obtained from MP2-SYN and MP2-UNI images were then warped to WMn-MRAGE image using the affine matrix obtained in the prior step. Visual inspection was performed to assess the quality of the automatic segmentations obtained from the different contrasts. Segmentation accuracy for each thalamic nucleus was assessed using the dice coefficients, volume similarity index (VSI) and distance between centroids of either the MP2-SYN or MP2-UNI derived segmentations and the corresponding segmentation from WMn-MPRAGE for all 13 subjects.

$$\text{dice}(A_i, B_i) = \frac{2|A_i \cap B_i|}{|A_i \cup B_i|} \text{ and } \text{VSI}(A, B) = 1 - \frac{(||A_i| - |B_i||)}{|A_i| + |B_i|}$$

For any nucleus $i$, the dice coefficient measured the overlap between the segmentation obtained from either MP2-SYN or MP2-UNI ($A_i$) and the segmentation obtained from WMn-MPRAGE image ($B_i$). Dice coefficient of 1 indicates perfect overlap between the segmentations. The volumetric similarity index (VSI) for a nucleus $i$ is ratio of the absolute volume difference and sum of the volume obtained from either MP2-SYN or MP2-UNI ($A_i$) and the volume obtained from WMn-MPRAGE image ($B_i$). A VSI of 1 indicates that the two nuclei have the same volume. The distance between centroids measures how close a thalamic nucleus obtained from the MP2-SYN or MP2-UNI is to the corresponding nucleus obtained from WMn-MPRAGE.

A paired t-test was used in Microsoft Excel for all comparisons, with multiple comparisons (12 nuclei plus whole thalamus = 13) accounted for using a Bonferroni corrected p-value threshold of $0.05/13 = 0.0038$.

RESULTS

The time taken to segment the thalamic nuclei for a single subject was approximately 30 minutes on a 6-core 3.5 GHz Intel Xeon MacPro. Table 1 shows the mean dice coefficients, volume similarity indices and centroid distance for the thalamic nuclei segmented using the two different THOMAS pipelines – MP2-SYN using THOMAS joint label fusion and MP2-UNI using THOMAS majority voting. WMn-MPRAGE using THOMAS joint label fusion was used as a reference. The data for the right and left nuclei has been aggregated as neither the dice coefficients nor the VSI or centroid distances of the nuclei were significantly different ($p > 0.05$) between the two hemispheres.

Table 1. Comparison of mean dice, VSI, and centroid distance for MP2-UNI and MP2-SYN (synthesized WMn) data using the WMn-MPRAGE as reference for the whole thalamus and 12 nuclei. The nuclei are sorted in decreasing order of bilateral volumes.

| | Volume (mm$^3$) | MP2-UNI Dice | MP2-SYN Dice | MP2-UNI VSI | MP2-SYN VSI | MP2-UNI Cent. dist | MP2-SYN Cent. dist |
|---|---|---|---|---|---|---|---|
| Thalamus | 10802 (1030) | 0.92 (0.02) | **0.94 (0.01)**$^*$ | 0.96 (0.02) | **0.99 (0.01)**$^*$ | 0.74 (0.28) | **0.33 (0.14)**$^*$ |
| Pul | 2754 (388) | 0.86 (0.03) | 0.90 (0.01)$^§$ | 0.95 (0.02) | **0.98 (0.01)**$^*$ | 0.83 (0.29) | **0.44 (0.08)**$^*$ |
| VLP | 1512 (175) | 0.84 (0.05) | **0.89 (0.02)**$^*$ | 0.97 (0.02) | 0.98 (0.02) | 0.82 (0.37) | **0.48 (0.18)**$^*$ |
| MD | 1288 (129) | 0.84 (0.04) | **0.89 (0.02)**$^*$ | 0.96 (0.03) | 0.98 (0.02)$^§$ | 1.01 (0.33) | **0.46 (0.21)**$^*$ |
| VA | 530 (75) | 0.85 (0.01) | 0.85 (0.02) | 0.98 (0.01) | 0.98 (0.01) | 0.37 (0.08) | 0.31 (0.09) |
| VPL | 517 (63) | 0.77 (0.05) | **0.85 (0.02)**$^*$ | 0.97 (0.02) | 0.97 (0.02) | 0.85 (0.33) | **0.42 (0.14)**$^*$ |
| AV | 235 (47) | 0.56 (0.15) | **0.69 (0.10)**$^*$ | 0.87 (0.06) | 0.85 (0.08) | 1.55 (0.63) | **0.78 (0.68)**$^*$ |
| CM | 207 (41) | 0.74 (0.06) | **0.82 (0.05)**$^*$ | 0.94 (0.06) | 0.95 (0.02) | 0.83 (0.30) | **0.50 (0.18)**$^*$ |
| LGN | 189 (28) | 0.71 (0.10) | 0.74 (0.08) | 0.93 (0.03)$^§$ | 0.86 (0.09) | 1.04 (0.63) | 0.55 (0.18)$^§$ |
| VLa | 135 (26) | **0.73 (0.06)**$^*$ | 0.66 (0.09) | 0.90 (0.06)$^§$ | 0.80 (0.10) | 0.69 (0.34) | 0.65 (0.31) |
| MGN | 125 (14) | 0.80 (0.08) | 0.84 (0.03) | 0.95 (0.03) | 0.94 (0.02) | 0.44 (0.24) | 0.34 (0.12) |
| MTT | 65 (15) | 0.59 (0.08) | 0.66 (0.03)$^§$ | 0.92 (0.04) | 0.93 (0.04) | 1.48 (0.50) | 0.94 (0.39)$^§$ |
| Hb | 41 (13) | 0.61 (0.11) | 0.61 (0.09) | 0.87 (0.09)$^§$ | 0.77 (0.11) | 0.82 (0.30) | 0.59 (0.17)$^§$ |

$^*$*Bonferroni corrected p<0.05/13=0.0038, $^§$p<0.05, SD in parentheses, bold indicates the larger Dice and VSI or smaller centroid distance for significantly different nuclei at p<0.0038. Pulvinar (Pul), Ventral Lateral Posterior (VLP), Mediodorsal (MD), Ventral Anterior (VA), Ventral Posterior Lateral (VPL), Anteroventral (AV), Center Median nucleus (CM), Lateral Geniculate Nucleus (LGN), Ventral Lateral anterior (VLa), Medial Geniculate Nucleus (MGN), Mammillothalamic Tract (MTT), Habenula (Hb).*

For the whole thalamus, the dice coefficients were 0.94 for the MP2-SYN and 0.92 for the MP2-UNI. Specifically, the five larger nuclei (volume 500 mm$^3$ or higher) achieved dice coefficients of > 0.85 and > 0.77 for MP2-SYN and MP2-UNI respectively. The dice coefficients obtained from MP2-SYN were significantly higher (p<0.05/13=0.038) than those obtained from MP2-UNI for the whole thalamus and three of the larger nuclei (VLp, MD, VPl) and two of the smaller nuclei (AV, CM) while VLa was the only nucleus for which dice coefficients from MP2-SYN was significantly lower than MP2-UNI. Similar patterns were observed for the VSI with higher VSIs for the nuclei segmented using MP2-SYN than MP2-UNI with the whole thalamus and Pul being significantly higher for MP2-SYN compared to MP2-UNI. The centroids of the MP2-SYN method were much closer to the WMn-MPRAGE reference than MP2-UNI.

The distance between centroids was less than 0.5 mm for the larger nuclei and less than 0.78 mm for the smaller nuclei, further attesting to the accuracy and reduced bias when using THOMAS on MP2-SYN.

Figure 3 shows the segmentation results from a single healthy subject using THOMAS on MP2-UNI (top two rows) and MP2-SYN (bottom two rows) respectively as colored overlays on the respective source images in axial, sagittal, and coronal planes. The same images without the overlays are shown in rows 1 and 3 for visualization of intra-thalamic contrast. The segmentation from WMn-MPRAGE (reference standard) is overlaid as outlines. Visually, the segmentation from MP2-UNI looks comparable to the MP2-SYN and both comport well with the ground truth outlines. This concordance reflects the high dice coefficients of Table 1.

Out of the 5 MS patients, 3 had visible intrathalamic lesions. Figure 4 shows the segmentation results from a MS patient using THOMAS on WMn-MPRAGE, MP2-SYN and MP2-UNI (columns from left to right). The THOMAS segmentation is shown as colored outline in the bottom row. Note the excellent segmentation achieved by THOMAS despite the presence of a fairly large lesion (arrow).

DISCUSSION

To our knowledge, this is the first study reporting segmentation of thalamic nuclei on 7T MP2RAGE data. The THOMAS algorithm successfully segmented thalamic nuclei in MP2RAGE images when evaluated against WMn-MPRAGE for all 13 subjects, even in the presence of MS lesions in some subjects. The high dice coefficients and high-volume similarity indices (Table 1) indicate that THOMAS works as well with ME-MP2RAGE as with the WMn-MPRAGE images. MP2-SYN achieved significantly higher dice coefficients in 4 large and 2 small nuclei compared to MP2-UNI attesting to the superior intra-thalamic contrast of white-matter nulling and possibly from the use of the more accurate joint fusion algorithm compared to majority voting (38). While not as high as MP2-SYN, excellent performance was achieved also with the MP2-UNI data. This may in part be due to the fact that the $T_1$ weighted image reconstructed from the ME-MP2RAGE sequence is a ratio image and has improved $B_1+$ uniformity as the transmit and receive coil heterogeneity effects are uniform across the different TIs and hence are nullified.

In this study, we have optimized and collected data using multi-echo MP2RAGE sequence. The advantage of ME-MP2RAGE over single echo MP2RAGE is that ME-MP2RAGE k-space data can be reconstructed to obtain T2* and quantitative susceptibility maps (35,39,40). The T1 values obtained with ME-MP2RAGE and the single echo MP2RAGE sequence show excellent agreement across brain regions and the T2* and susceptibility weighted imaging (SWI) maps obtained from ME-MP2RAGE data were of comparable quality to those obtained from Fast Low-Angle SHot (FLASH) data (35,39,40). Hence a single ME-MP2RAGE sequence can yield multiple quantitative maps of tissue relaxometry in a favorable acquisition time (without sacrificing image quality) over multiple separate sequences for T1, T2* and SWI maps.

The library of 20 priors in the THOMAS atlas (34) included 11 datasets from adult MS patients. In order to accurately detect subtle changes in volume between groups, the composition of the priors should be a representation of the population being studied. This is especially true when characterizing focal changes in volume in MS where thalamic atrophy and ventricular enlargement are both hallmarks of the disease. Of relevance to studies in adolescents and young adults, the thalamic volumes reach maximal values in early adolescence and have a relatively flat growth trajectory into early adulthood, and thus an atlas that included adult MS participants is appropriate for the age group studied in the present work (5).

Recently, a probabilistic atlas constructed from manual delineation of 26 thalamic nuclei on six autopsy specimens combined with Bayesian inference methods has been used to segment 3T MPRAGE images into 26 nuclei per side (41). However, this method is very time consuming (6 to 10 hours) as it requires the whole Freesurfer pipeline to be run beforehand and has neither been validated against manual segmentation on in-vivo images nor on 7T images.

Another recent study reported manual identification of 23 nuclei on 9 healthy subjects on multiple 7T MPRAGE images (with different inversion times) and SWI. The segmentations were then used to train shape models that were used to segment thalamic nuclei on 3T MPRAGE images (42). Models based solely on shape priors from healthy subjects can be useful in certain applications. However, in disease, where different nuclei are impacted and possibly atrophied nonuniformly resulting in shape changes (6), shape-based models trained only on healthy subjects may not be able to segment those nuclei accurately. Indeed, in this study, some smaller nuclei such as the lateral geniculate nuclei (LGN) that vary multifold in shape and size (43), were not included.

Our study had a few limitations including a small subject size (n=13) and lack of manual segmentation gold standard, which forced the use of WMn-MPRAGE segmentation as a "silver" standard. THOMAS previously has been rigorously evaluated on WMn-MPRAGE at 7T by comparison to manual segmentation and has been shown to be highly accurate (>0.85 dice for larger nuclei and > 0.7 dice for smaller nuclei) (Su 2019). In the context of segmentation validation, a Dice value of 0.7 or greater provides good overlap (44,45) and gives high correlation with Probabilistic Distance, a metric that is highly sensitive to alignment errors (46). A Dice > 0.7 also gives high correlation with Volumetric Similarity (VS) or AVerage Distance metrics, that are important when volume or position of the contour are of greatest interest, respectively. The MEMP2RAGE data was at 0.66 mm isotropic resolution whilst WMn-MPRAGE was at 1mm isotropic resolution, which could have made the dice coefficients suboptimal due to interpolation errors. However, studies using THOMAS will compute thalamic nuclei volumes in their native resolution without interpolation, thus mitigating such errors.

CONCLUSIONS

THOMAS can reliably segment thalamic nuclei using MP2RAGE images. With MP2RAGE sequence increasingly being used in 7T studies, the successful demonstration of thalamic nuclei delineation suggests wider applicability of our method to more studies of thalamic involvement in aging and diseases.

ACKNOWLEDGEMENTS:

The authors would like to thank Jackie Meekes, Deepa Thakuri and Candice Dunn for technical assistance. Study Funded by the National Multiple Sclerosis Society (NMSS PP-1901-33149 and NIBIB - 3P41EB015893)

FIGURES:

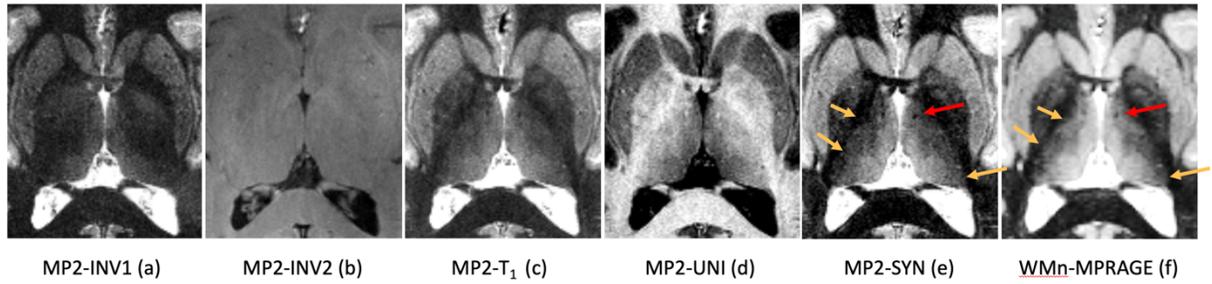

**Figure 1** Different contrasts obtained from ME-MP2RAGE and WMn-MPRAGE sequences. (a-b) images obtained from the first and second inversion times in the ME-MP2RAGE sequence; (c-e) quantitative $T_1$ map, bias corrected $T_1$-weighted (MP2-UNI) image and white-matter-nulled image synthesized (MP2-SYN) from the $T_1$ map obtained from ME-MP2RAGE sequence; (f) white-matter-nulled image from WMn-MPRAGE sequence.

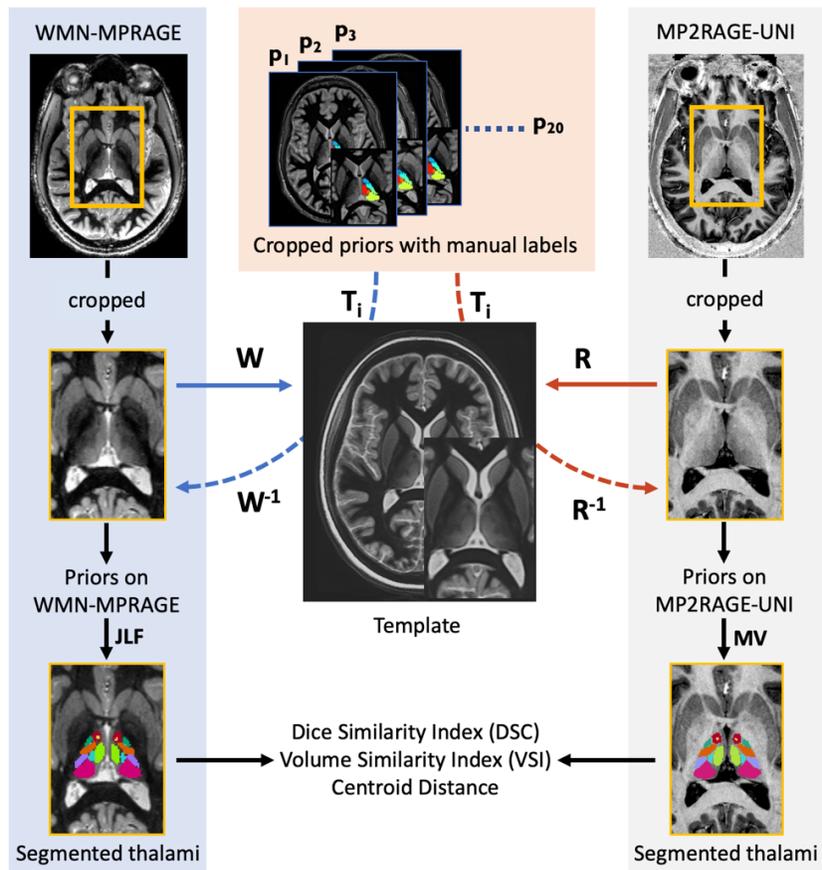

**Figure 2** Default and modified THOMAS workflow. The left column shows the default settings in the THOMAS pipeline that uses joint label fusion (JLF) to process and segment thalamic nuclei from white-matter-nulled contrast synthesized from ME-MP2RAGE sequence and the true WMn-MPRAGE images. The right column shows a modification of the THOMAS pipeline on MP2-UNI image wherein majority voting algorithm is used instead of joint label fusion.

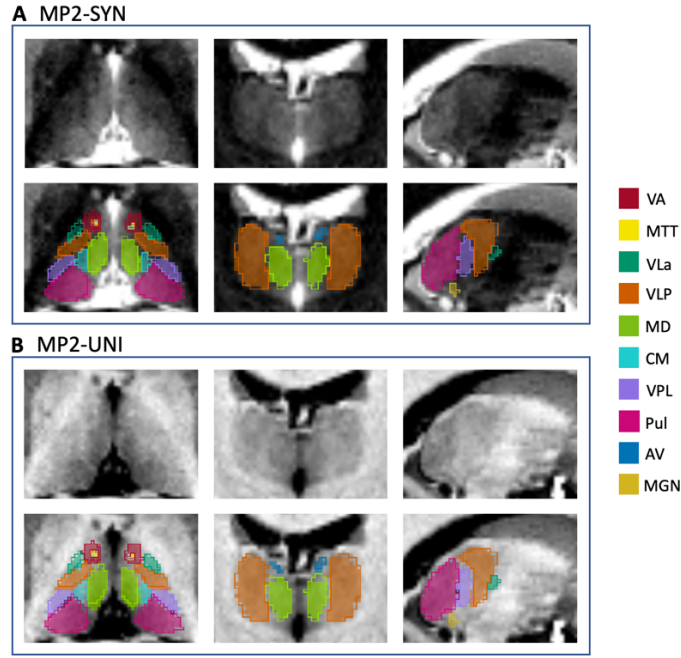

**Figure 3** Automatic segmentation of different thalamic nuclei overlaid on a healthy subject's WMn-MPRAGE. In the middle row, the filled-in solid overlay correspond to thalamic segmentation obtained from MP2-UNI and the overlay in outline are the segmentations obtained from WMn-MPRAGE (silver standard). In the bottom row, the filled-in solid overlay correspond to thalamic segmentation obtained from MP2-SYN and the overlay in outline are the segmentations obtained from WMn-MPRAGE (silver standard). Refer to Table 1 legend for thalamic nuclei abbreviations

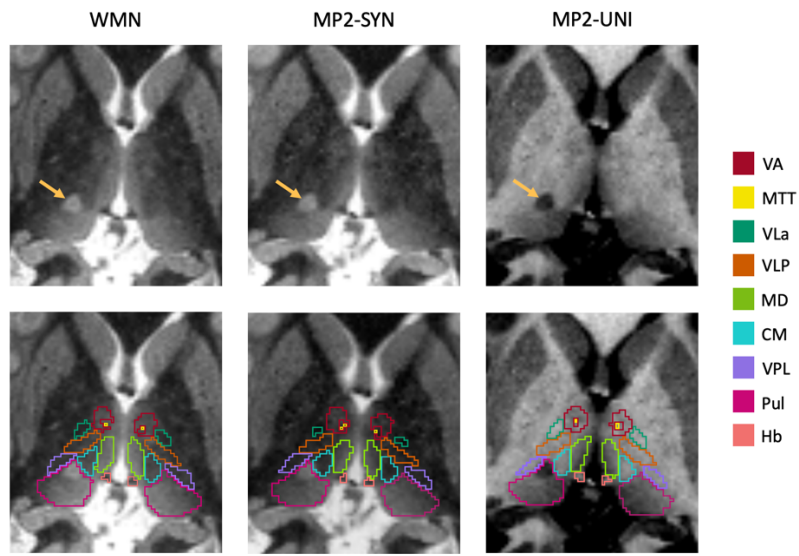

**Figure 4** shows the segmentation results from a subject with pediatric-onset MS using THOMAS on WMn-MPRAGE, MP2-SYN and MP2-UNI (in the columns from left to right). The location of the lesion is demarcated by arrow in the top row and the THOMAS segmentation is shown as colored outline in the bottom row. Note the excellent segmentation achieved by THOMAS despite the presence of a lesion in the data. Refer to Table 1 legend for thalamic nuclei abbreviations

REFERENCES:

1. Morel A, Magnin M, Jeanmonod D. Multiarchitectonic and stereotactic atlas f the human thalamus. J. Comp. Neurol. 1997 doi: 10.1002/(SICI)1096-9861(19971103)387:4<588::AID-CNE8>3.0.CO;2-Z.
2. Krauth A, Blanc R, Poveda A, Jeanmonod D, Morel A, Székely G. A mean three-dimensional atlas of the human thalamus: Generation from multiple histological data. Neuroimage 2010 doi: 10.1016/j.neuroimage.2009.10.042.
3. Sherman SM, Guillery RW. The role of the thalamus in the flow of information to the cortex. In: Philosophical Transactions of the Royal Society B: Biological Sciences. ; 2002. doi: 10.1098/rstb.2002.1161.
4. Theyel BB, Llano DA, Sherman SM. The corticothalamocortical circuit drives higher-order cortex in the mouse. Nat. Neurosci. 2010 doi: 10.1038/nn.2449.
5. Aubert-Broche B, Fonov V, Narayanan S, et al. Onset of multiple sclerosis before adulthood leads to failure of age-expected brain growth. Neurology 2014 doi: 10.1212/WNL.0000000000001045.
6. Bisecco A, Capuano R, Caiazzo G, et al. Regional changes in thalamic shape and volume are related to cognitive performance in multiple sclerosis. Mult. Scler. J. 2019 doi: 10.1177/1352458519892552.
7. Rojas JI, Murphy G, Sanchez F, et al. Thalamus volume change and cognitive impairment in early relapsing–remitting multiple sclerosis patients. Neuroradiol. J. 2018 doi: 10.1177/1971400918781977.
8. Younce JR, Campbell MC, Perlmutter JS, Norris SA. Thalamic and ventricular volumes predict motor response to deep brain stimulation for Parkinson's disease. Park. Relat. Disord. 2019 doi: 10.1016/j.parkreldis.2018.11.026.
9. Aggleton JP, Pralus A, Nelson AJD, Hornberger M. Thalamic pathology and memory loss in early Alzheimer's disease: Moving the focus from the medial temporal lobe to Papez circuit. Brain 2016 doi: 10.1093/brain/aww083.
10. Chand GB, Dwyer DB, Erus G, et al. Two distinct neuroanatomical subtypes of schizophrenia revealed using machine learning. Brain 2020 doi: 10.1093/brain/awaa025.
11. Chye Y, Mackey S, Gutman BA, et al. Subcortical surface morphometry in substance dependence: An ENIGMA addiction working group study. Addict. Biol. 2019 doi: 10.1111/adb.12830.
12. Segobin S, Laniepce A, Ritz L, et al. Dissociating thalamic alterations in alcohol use disorder defines specificity of Korsakoff's syndrome. Brain 2019 doi: 10.1093/brain/awz056.
13. Chen XY, Chen ZY, Dong Z, Liu MQ, Yu SY. Regional volume changes of the brain in migraine chronification. Neural Regen. Res. 2020 doi: 10.4103/1673-5374.276360.
14. Magon S, Tsagkas C, Gaetano L, et al. Volume loss in the deep gray matter and thalamic subnuclei: a longitudinal study on disability progression in multiple sclerosis. J. Neurol. 2020 doi: 10.1007/s00415-020-09740-4.
15. Planche V, Su JH, Mournet S, et al. White-matter-nulled MPRAGE at 7T reveals thalamic lesions and atrophy of specific thalamic nuclei in multiple sclerosis. Mult. Scler. J. 2019 doi: 10.1177/1352458519828297.
16. Low A, Mak E, Malpetti M, et al. Asymmetrical atrophy of thalamic subnuclei in Alzheimer's disease and amyloid-positive mild cognitive impairment is associated with key clinical features. Alzheimer's Dement. Diagnosis, Assess. Dis. Monit. 2019 doi: 10.1016/j.dadm.2019.08.001.
17. Vassal F, Coste J, Derost P, et al. Direct stereotactic targeting of the ventrointermediate nucleus of the thalamus based on anatomic 1.5-T MRI mapping with a white matter attenuated inversion recovery (WAIR) sequence. Brain Stimul. 2012 doi: 10.1016/j.brs.2011.10.007.
18. Boutet A, Ranjan M, Zhong J, et al. Focused ultrasound thalamotomy location determines clinical benefits in patients with essential tremor. Brain 2018 doi: 10.1093/brain/awy278.
19. Behrens TEJ, Johansen-Berg H, Woolrich MW, et al. Non-invasive mapping of connections between human thalamus and cortex using diffusion imaging. Nat. Neurosci. 2003 doi: 10.1038/nn1075.
20. Fan Y, Nickerson LD, Li H, et al. Functional Connectivity-Based Parcellation of the Thalamus: An Unsupervised Clustering Method and Its Validity Investigation. Brain Connect. 2015 doi: 10.1089/brain.2015.0338.

21. Glaister J, Carass A, Stough J V., Calabresi PA, Prince JL. Thalamus parcellation using multi-modal feature classification and thalamic nuclei priors. In: Medical Imaging 2016: Image Processing. ; 2016. doi: 10.1117/12.2216987.
22. Duan Y, Li X, Xi Y. Thalamus segmentation from diffusion tensor magnetic resonance imaging. Int. J. Biomed. Imaging 2007 doi: 10.1155/2007/90216.
23. Mang SC, Busza A, Reiterer S, Grodd W, Klose AU. Thalamus segmentation based on the local diffusion direction: A group study. Magn. Reson. Med. 2012 doi: 10.1002/mrm.22996.
24. Kumar V, Mang S, Grodd W. Direct diffusion-based parcellation of the human thalamus. Brain Struct. Funct. 2015 doi: 10.1007/s00429-014-0748-2.
25. Lambert C, Simon H, Colman J, Barrick TR. Defining thalamic nuclei and topographic connectivity gradients in vivo. Neuroimage 2017 doi: 10.1016/j.neuroimage.2016.08.028.
26. Najdenovska E, Alemán-Gómez Y, Battistella G, et al. In-vivo probabilistic atlas of human thalamic nuclei based on diffusion-weighted magnetic resonance imaging. Sci. Data 2018 doi: 10.1038/sdata.2018.270.
27. Battistella G, Najdenovska E, Maeder P, et al. Robust thalamic nuclei segmentation method based on local diffusion magnetic resonance properties. Brain Struct. Funct. 2017 doi: 10.1007/s00429-016-1336-4.
28. van Oort ESB, Mennes M, Navarro Schröder T, et al. Functional parcellation using time courses of instantaneous connectivity. Neuroimage 2018 doi: 10.1016/j.neuroimage.2017.07.027.
29. Marques JP, Kober T, Krueger G, van der Zwaag W, Van de Moortele PF, Gruetter R. MP2RAGE, a self bias-field corrected sequence for improved segmentation and T1-mapping at high field. Neuroimage 2010 doi: 10.1016/j.neuroimage.2009.10.002.
30. Marques JP, Gruetter R. New Developments and Applications of the MP2RAGE Sequence - Focusing the Contrast and High Spatial Resolution R1 Mapping. PLoS One 2013 doi: 10.1371/journal.pone.0069294.
31. Okubo G, Okada T, Yamamoto A, et al. MP2RAGE for deep gray matter measurement of the brain: A comparative study with MPRAGE. J. Magn. Reson. Imaging 2016 doi: 10.1002/jmri.24960.
32. Tourdias T, Saranathan M, Levesque IR, Su J, Rutt BK. Visualization of intra-thalamic nuclei with optimized white-matter-nulled MPRAGE at 7T. Neuroimage 2014 doi: 10.1016/j.neuroimage.2013.08.069.
33. Saranathan M, Tourdias T, Bayram E, Ghanouni P, Rutt BK. Optimization of white-matter-nulled magnetization prepared rapid gradient echo (MP-RAGE) imaging. Magn. Reson. Med. 2015 doi: 10.1002/mrm.25298.
34. Su JH, Thomas FT, Kasoff WS, et al. Thalamus Optimized Multi Atlas Segmentation (THOMAS): fast, fully automated segmentation of thalamic nuclei from structural MRI. Neuroimage 2019 doi: 10.1016/j.neuroimage.2019.03.021.
35. Metere R, Kober T, Möller HE, Schäfer A. Simultaneous quantitative MRI mapping of T1, T2∗ and magnetic susceptibility with Multi-Echo MP2RAGE. PLoS One 2017 doi: 10.1371/journal.pone.0169265.
36. Thompson AJ, Banwell BL, Barkhof F, et al. Diagnosis of multiple sclerosis: 2017 revisions of the McDonald criteriaTHOMPSON, A. J. et al. Diagnosis of multiple sclerosis: 2017 revisions of the McDonald criteria. The Lancet Neurology, v. 17, n. 2, p. 162–173, 2018. Lancet Neurol. 2018;17:162–173 doi: 10.1016/S1474-4422(17)30470-2.
37. Avants BB, Tustison NJ, Song G, Cook PA, Klein A, Gee JC. A reproducible evaluation of ANTs similarity metric performance in brain image registration. Neuroimage 2011 doi: 10.1016/j.neuroimage.2010.09.025.
38. Wang H, Yushkevich PA. Multi-atlas segmentation with joint label fusion and corrective learning—an open source implementation. Front. Neuroinform. 2013 doi: 10.3389/fninf.2013.00027.
39. Sun H, Cleary JO, Glarin R, et al. Extracting more for less: multi-echo MP2RAGE for simultaneous T 1 -weighted imaging, T 1 mapping, mapping, SWI, and QSM from a single acquisition. Magn. Reson. Med. 2020;83:1178–1191 doi: 10.1002/mrm.27975.
40. Caan MWA, Bazin PL, Marques JP, de Hollander G, Dumoulin SO, van der Zwaag W. MP2RAGEME: T 1 , T 2* , and QSM mapping in one sequence at 7 tesla. Hum. Brain Mapp. 2019 doi: 10.1002/hbm.24490.
41. Iglesias JE, Insausti R, Lerma-Usabiaga G, et al. A probabilistic atlas of the human thalamic nuclei combining ex vivo MRI and histology. Neuroimage 2018 doi: 10.1016/j.neuroimage.2018.08.012.

42. Liu Y, D'Haese PF, Newton AT, Dawant BM. Generation of human thalamus atlases from 7 T data and application to intrathalamic nuclei segmentation in clinical 3 T T1-weighted images. Magn. Reson. Imaging 2020 doi: 10.1016/j.mri.2019.09.004.

43. Li M, He H, Shi W, et al. Quantification of the human lateral geniculate nucleus in vivo using MR imaging based on morphometry: Volume loss with age. Am. J. Neuroradiol. 2012 doi: 10.3174/ajnr.A2884.

44. Zijdenbos AP, Dawant BM, Margolin RA, Palmer AC. Morphometric Analysis of White Matter Lesions in MR Images: Method and Validation. IEEE Trans. Med. Imaging 1994 doi: 10.1109/42.363096.

45. Zou KH, Warfield SK, Bharatha A, et al. Statistical Validation of Image Segmentation Quality Based on a Spatial Overlap Index. Acad. Radiol. 2004 doi: 10.1016/S1076-6332(03)00671-8.

46. Taha AA, Hanbury A. Metrics for evaluating 3D medical image segmentation: Analysis, selection, and tool. BMC Med. Imaging 2015 doi: 10.1186/s12880-015-0068-x.